\documentclass[12pt]{article}
\pdfoutput=1

\usepackage{amsmath,amssymb,amsbsy,amsfonts,latexsym,graphicx}
\usepackage{hyperref}
\usepackage{color,array,subfigure}
\usepackage{cite}

\newcommand{\m}{\mu}

\allowdisplaybreaks

\evensidemargin \oddsidemargin

 \def\m{{\mu}}

\begin{document}


\begin{titlepage}

\renewcommand{\thefootnote}{\fnsymbol{footnote}}



\vspace{15mm}
\baselineskip 9mm
\begin{center}
  {\Large \bf 
{  
Building Magnetic Hysteresis in Holography} 
 }
\end{center}

\baselineskip 6mm
\vspace{10mm}
\begin{center}
Kyung Kiu Kim$^{1,a}$, Keun-Young Kim$^{2,b}$, Yunseok Seo$^{3,c}$ and Sang-Jin Sin$^{4,d}$
 \\[10mm] 
  $^1${\sl Department of Physics and Astronomy, Sejong University, Seoul 05006, Korea}
   \\[3mm]
    $^2${\sl School of Physics and Chemistry, \\ Gwangju Institute of Science and Technology,  Gwangju 61005, Korea}
     \\[3mm]
    $^3${\sl GIST College, Gwangju Institute of Science and Technology,  Gwangju 61005, Korea}
  \\[3mm]
    $^4${\sl Department of Physics, Hanyang University, Seoul 133-791, Korea}
  \\[3mm]
  {\tt  ${}^a$kimkyungkiu@sejong.ac.kr,\,${}^b$fortoe@gist.ac.kr,
  \\${}^c$yseo@gist.ac.kr,\,${}^d$sjsin@hanyang.ac.kr}
\end{center}

\thispagestyle{empty}

\vspace{1cm}
\begin{center}
{\bf Abstract}
\end{center}
\noindent
We study the spontaneous magnetization and the magnetic hysteresis using the gauge/gravity duality. We first propose a novel and general formula to compute the magnetization in a large class of holographic models. By using this formula, we compute the spontaneous magnetization 
in a model like a holographic superconductor.
Furthermore, we turn on the external magnetic field and
build the hysteresis curve of magnetization and charge density.
To our knowledge, 
this is the first holographic model realizing  
the hysteresis accompanied with spontaneous symmetry breaking. 
\\ [15mm]

\vspace{5mm}
\end{titlepage}

\baselineskip 6.6mm
\renewcommand{\thefootnote}{\arabic{footnote}}
\setcounter{footnote}{0}

\section{Introduction}

The magnetism is one of the most intriguing topics in various fields of physics. One of the reasons is that interesting subjects of the magnetism are closely related to a fundamental quantum degrees of freedom that distinguish kinds of particles, the particle's statistics and phases of the system. One of the significant phenomena of magnetic systems is the ferromagnetism given by the spontaneous symmetry breaking. In addition, the ferromagnetic system usually shows hysteresis curves, which are representative phenomena of the ferromagnetism. As a preliminary study toward ferromagnetism, we will consider a simple model to build spontaneous magnetization and the hysteresis curve using gauge/gravity duality \cite{Maldacena:1997re,Aharony:1999ti} . In earlier studies, magnetized systems have already attracted much attention and have been studied numerous times in the holographic approach, {\it e.g,}\cite{Hartnoll:2007ai,Hartnoll:2007ih,Albash:2008eh,Jensen:2011xb,Kim:2013wiz,Blake:2014yla,Donos:2012yu,Seo:2015pug,Blake:2015epa,Kim:2015wba,Blake:2015hxa,Blake:2015ina,Cai:2015bsa,Cai:2015xpa,Cai:2015jta,Donos:2017mhp}.

The gauge/gravity duality is a very powerful tool to study strongly coupled field theory systems. The main advantage of the duality is that the generating functional of the strongly coupled field theory can be obtained by the on-shell action of the corresponding gravity theory \cite{Gubser:1998bc,Witten:1998qj}. This correspondence has been applied to various area of physics. One of the active applications is about condensed matter theories, {\it e.g,}\cite{Hartnoll:2007ih,Herzog:2007ij,Herzog:2008he,Hartnoll:2008vx,Hartnoll:2008kx,Gubser:2009qm,Herzog:2009xv}.

More specifically, the gauge/gravity duality is a relation between a $d$-dimensional field theory living on the boundary and the corresponding $(d+1)$-dimensional bulk gravitational system with an appropriate boundary condition. Such a boundary condition tells us the prescription for the external sources and the non-normalizable modes of the field theory and the gravitational system, respectively. In this subject, we mainly consider a system with the sources given by the external magnetic field and chemical potential. 
These sources induce the magnetization, charge density and vacuum expectation value(VEV) of a scalar operator. We take into account $(2+1)$-dimensional systems in the present work.

The $(2+1)$-dimensional spacetime we considered has a very interesting magnetic nature because the direction of magnetic field is always orthogonal to the $2$-dimensional space when the spacetime is embedded in higher dimensions. The magnetization is described as a real value, just as the magnetic field is. Therefore, the spontaneous magnetization is displayed by breaking $\mathbb{Z}_2$ symmetry. To realize this spontaneous symmetry breaking, we use the well-known hairy black brane solution with a real scalar field $\phi$. Such a real scalar field is dual to a scalar operator in the boundary theory. Since our final goal is to construct spontaneous magnetization and hysteresis curves, we introduce an interaction among the scalar operator, the magnetic field and the charge density operator. In bulk point of view, this interaction is achieved by adding an axionic term, $\int d^4 x W(\phi)\epsilon^{MNPQ}F_{MN}F_{PQ}$,\, to the action. And we consider a large class of model which can describe the spontaneous magnetization. We propose a general formula for the magnetization by using a `scaling symmetry' trick \cite{Banados:2005hm,Ahn:2015shg,Ahn:2015uza}. This formula can be confirmed in a specific model. 
The proposed large class of models enable us to describe the spontaneous magnetization.

As we mentioned, one of the characteristics of magnetic material is the hysteresis curve of magnetization. In order to construct the hysteresis curves, the interaction potential $W(\phi)$ plays an important role. This is because the interaction potential $W(\phi)$ should be an odd function to break the $\mathbb{Z}_2$ symmetry    changing the sign of the scalar field $\phi$. We will consider an odd function for a specific model. Also, in fact, the hysteresis is obtained by the slowly varying magnetic field. In the real physics, we must consider a time-dependent background.  Since, however, we assume a very slowly varying magnetic field, we ignore the effect from the time dependent dynamics.
See \cite{Evans:2010xs} for an explicit time-dependent consideration for another holographic model.

This paper is organized as follows: In section \ref{sec2}, we propose the general form of the magnetization using the `scaling symmetry' trick and introduce a simple model. Then we check that our general formula works in the specific model. In addition, we find the expression of the on-shell action dual to the free energy in the boundary field theory. In section \ref{sec3}, we investigate the spontaneous magnetization and the hysteresis curve. 
In section \ref{sec5}, we discuss our result and introduce possible future directions. Several formulas related to the holographic renormalization and a description in Landau-Ginzburg type potential are summarized in Appendix.

\section{Basic Formulation}\label{sec2}

In this section, we consider a large class of holographic models describing spontaneous magnetization. The magnetization is given by derivative of free energy with respect to external magnetic field $\mathcal{B}$ as follows\footnote{See \cite{Hartnoll:2007ih} for details.}:
\begin{align}\label{mag01}
\mathcal M = - \frac{\delta \Omega}{\delta \mathcal{B}}~~,
\end{align}
where $\Omega$ is the free energy of $d$-dimensional system. This is the most essential quantity to understand physics in various magnetic materials. The main goal of the present work is to study this quantity through holographic approach.

The fundamental relation of thermodynamics is the first law describing a constraint among variations of thermodynamic quantities. The gauge/gravity duality enables us to interpret this relation in the bulk geometry. $(d+1)$-dimensional black holes also have relations among their parameters by on-shell variation. These two relations have a similarity, which leads to the holographic principle and the gauge/gravity correspondence. Especially, we are interested in the magnetic variation. It consists of the external magnetic field variation and the magnetization. The form of 4-dimensional magnetic black brane solutions is simpler than those in the higher dimensions. Thus we will focus on the asymptotically $AdS_4$ black branes in this paper.

Furthermore, this first law of thermodynamics can be integrated to a relation among finite thermodynamic parameters. The resulting relation is consistent with the definition of free energy density by construction. One may write down the relation as follows: 
\begin{align}\label{pressure}
\omega\equiv\Omega/\mathcal{V} 
= \epsilon - s T  - \mu\, \tilde{Q} ~~,
\end{align}
where $\omega$ is the free energy density of the system. In addition, spatial volume, energy density, entropy density, temperature, chemical potential and charge density are denoted by $\mathcal{V}$, $\epsilon$, $s$, $T$, $\mu$, $\tilde Q$, respectively. Also, it is well known that the pressure $\mathcal P$ is same with $-\,\omega$ for homogeneous systems.

When the system has the magnetization $\mathcal M$ under the magnetic field $\mathcal B$, the pressure consists of the magnetic part and the other contribution as follows:
\begin{align}\label{P+Pm}
\mathcal P = \mathcal{P}_{others} + \mathcal M\,\mathcal{B}~~.
\end{align}
From this decomposition one can easily read off the magnetization $\mathcal{M}$. In the bulk geometry, the above relation (\ref{pressure}) corresponds to the so-called Smarr relation. It is the constraint among black hole parameters given by vanishing time-time component of the metric at the horizon. We will show this explicitly.

Recently the general form of the Smarr relation has been studied and was derived by applying a modified Noether theorem to a reduced action of gravitational systems \cite{Banados:2005hm,Hyun:2015tia,Ahn:2015uza,Ahn:2015shg,Hyun:2016isn,Hyun:2017nkb}. This is very useful even in a hairy black hole background. We will extend the method for hairy magnetic black branes.

\subsection{Einstein-Maxwell-Dilaton Model}

Let us start with a large class of holographic models to describe magnetized systems.
\begin{align}\label{Action00}
S = &\,\frac{1}{16\pi G}\int  d^4 x \sqrt{-g}  \left\{   R + \frac{6}{L^2}    -\frac{Z(\phi)}{4} F^2   - \frac{1}{  2} \left(\partial \phi\right)^2   - V(\phi)     \right\} \nonumber\\&\,+ \frac{\theta}{16\pi G}   \int  d^4 x W(\phi) \epsilon^{MNPQ} F_{MN} F_{PQ} ~~, 
\end{align}
where the last term plays an important role for the spontaneous magnetization and the hysteresis curve. In 3+1 dimension, this term is linear in the magnetic field. So one may notice that there can exist a non-vanishing magnetization even in the absence of magnetic field. See (\ref{mag01}) for this speculation. Furthermore, turning on the magnetic field with even functions of $Z(\phi)$ and $V(\phi)$, 
an odd function of $W(\phi)$ breaks a $\mathbb{Z}_2$ symmetry by changing the sign of the scalar field($\phi \to - \phi$). In this case, however, the Lagrangian is invariant under another transformation that simultaneously changes the sign of the magnetic field ($\mathcal{B}\to - \mathcal{B}$). Since the latter symmetry is essential to build the hysteresis curve, we will take into account the odd function $W(\phi)$ with even $Z(\phi)$ and $V(\phi)$ in a specific calculation. It will be clear in section \ref{sec3}.

Now, we will derive the Smarr relation of black branes in the gravity model using a `scaling symmetry' method in \cite{Banados:2005hm,Ahn:2015uza,Ahn:2015shg,Hyun:2016isn,Hyun:2017nkb}. In order to apply the method, we take the following metric ansatz:
\begin{align}\label{ansatz0}
&ds^2 = - U(r)e^{-\mathcal{W}(r)} d t^2  + \frac{r^2}{L^2} \left( dx^2+ dy^2 \right) + \frac{dr^2}{U(r)}~~,~~\\
&\phi=\phi(r)~~,~~A= A_{t}(r)d t + \frac{H}{2}(xdy-ydx)
\end{align}
We will use a convenient time coordinate $\tau = e^{-\mathcal{W}(\infty)/2}\, t$ to keep the boundary metric $\eta_{\mu\nu}$\,.  
As a result, the coordinate system of dual field theory becomes $(\tau,x,y)$ whose temporal component of the metric and gauge field is given by
\begin{align}
g_{\tau\tau}(r) = - U(r) e^{-(\mathcal{W}(r)-\mathcal{W}(\infty)) } ~~,~~A_\tau(r) = A_t(r)e^{\mathcal{W}(\infty)/2}~~.
\end{align} 
For simplicity, we define $\tilde{\mathcal{W}}(r)\equiv\mathcal{W}(r)-
\mathcal{W}(\infty)$. Then the Hawking temperature and entropy density are given by the following forms:
\begin{align}
T_H = \frac{U'(r_h)}{4\pi}e^{-\tilde{\mathcal W}(r_h)/2}~~,~~s = \frac{r_h^2}{4G L^2}~~,
\end{align}
where $r_h$ is the location of the horizon defined by $U(r_h)=0$.

Plugging our ansatz (\ref{ansatz0}) into the gravitational action (\ref{Action00}), we obtained the following reduced action: 
\begin{align}
S = \frac{1}{16\pi G} \int d\tau dx dy \int dr  L_{red}~~,
\end{align}
where
\begin{align}\label{Lred}
 L_{red}= &
\frac{e^{-\tilde{\mathcal{W}}(r)/2}}{L^2}
 \left(\frac{6 r^2}{L^2}-\frac{H^2 L^4 Z(\phi )}{2 r^2}-\frac{r^2 U(r) \phi'(r)^2}{2} -r^2 V({\phi}(r))-2 r U'(r)-2 U(r)\right) \nonumber\\
&+\frac{ r^2 Z(\phi ) e^{\frac{\tilde{\mathcal{W}}(r)}{2 }} A_{\tau }'(r){}^2}{2 L^2}-8 \,\theta  H\, W(\phi ) A_{\tau }'(r)~.
\end{align}
Here we dropped out total derivative terms. When $H=0$, the reduced action has a scaling symmetry. One can show that this reduced action without magnetic field is invariant under the following `scaling transformation' up to total derivatives. We defined the scaling transformation as follows:  
\begin{align}
&\delta( e^{-\tilde{\mathcal{W}}/2} )= - \lambda \left( 3\,e^{-\tilde{\mathcal{W}}/2} + r\,(e^{-\tilde{\mathcal{W}}/2})' \right)~~,~~ \\
&\delta U = -\lambda\,(-2\, U + r\, U')~~,\\
&\delta A_\tau = -\lambda\,( 2\, A_\tau + r A_\tau' )~~,\\
&\delta \phi = -\lambda\,r\,\phi'~~,
\end{align}
where $\lambda$ is a small parameter. On the other hand, this symmetry can be broken by turning on the magnetic field $H$. So we considered the invariant part of the Lagrangian. And one can show that 
\begin{align}\label{varL0}
\delta ( L_{red} - L_H ) = \lambda \left( - r ( L_{red} - L_H ) \right)',
\end{align}
where $L_H$ is the non-invariant part of the reduced Lagrangian (\ref{Lred}). It is defined as follows:
\begin{align}
L_H = -\left( 8\, \theta H W(\phi ) A_\tau'+\,e^{-\tilde{\mathcal W}/2}\frac{H^2 L^2 Z(\phi )}{2 r^2} \right)~~.
\end{align}
Using the equations of motion, $\delta L_{red}$ is given as follows:
\begin{align}\label{delLred}
\delta L_{red}= \left( \sum_i \delta \Psi_i \frac{\partial L_{red}}{\partial \Psi'_i }  \right)' ~~,
\end{align}
where $\Psi_i$ denotes a collective expression of the fields such as $\left\{e^{-\tilde{\mathcal{W}}/2}, U, A_\tau, \phi\right\}$. (\ref{varL0}) and (\ref{delLred}) can be incorporated to describe a partially conserved `charge'(PCC) $\mathcal C$.
\begin{align}\label{PCC}
\lambda\, \frac{d}{dr} \mathcal C  = \frac{1}{32\pi G} \left(\delta L_H + \lambda\,\left( r L_{H} \right)'\right)= -\frac{1}{32\pi G}\,\delta_H L_H ~~,
\end{align}
where $\mathcal C$ is defined by
\begin{align}
\mathcal C \equiv \frac{1}{32 \pi G\,\lambda}\left(\sum_i \delta \Psi_i \frac{\partial L_{red}}{\partial \Psi'_i } + \lambda\,r   L_{red}\right)~~.
\end{align}
and $\delta_H$ is the complementary transformation acting on $H$ to make the $L_{red}$ invariant under a fake `scaling transformation':
\begin{align}
\delta_H H = 2\,\lambda\, H~~.
\end{align}
The origin of the PCC equation (\ref{PCC}) is nothing but the Hamiltonian constraint along the radial direction with other equations of motion. Even though PCC does not introduce a new symmetry, it is very useful to describe thermodynamics of black holes.

Now, we can easily find the form of PCC as follows : 
\begin{align}
\mathcal C 
=-  A_\tau \mathcal{Q} +\frac{1}{32\pi G\,L^2} \,r\, e^{-\tilde{\mathcal{W}}/2} \left( 2 r\, U'+ r^2\,U\, {\phi'}^2 - 4\, U \right)~~,
\end{align}
where $\mathcal Q$ is the charge density of the black brane, which is  given by 
\begin{align}\label{chargeQ}
\mathcal Q = \frac{1}{16\pi G\,L^2 } \left( r^2  e^{\tilde{\mathcal{W}}/2} Z(\phi)A_\tau'(r)  - 8\, \theta\, W(\phi) H\,L^2\right)~.
\end{align}
We may take an integration of PCC over the outer horizon region. 
\begin{align}
\mathcal{C}(r) -\mathcal{C}(r_h) = -\frac{1}{32\pi G\,\lambda} \int_{r_h}^r dr' \delta_H L_H(r')~~,
\end{align}
where PCC at the horizon is given by $\mathcal{C}(r_h)=  s T_H  $ with adopting a regularity condition $A_{\tau}(r_h)=\,U(r_h)=0$.

Since the PCC is finite at the boundary of AdS space, it is legitimate to consider the PCC at $r=\infty$. Therefore, the integration over the outer region gives us the following relation:
\begin{align}
s T_H = \left(\mathcal{C}(\infty) + \frac{1}{32\pi\,G\,\lambda}\int_{r_H}^\infty dr'\delta_H L_H(r') \right)~~.
\end{align}
We will identify the temperature in the dual field theory $T$ with the Hawking temperature. In addition the asymptotic value of the PCC is given by 
\begin{align}
\mathcal{C}(\infty) = - \mu\, \tilde{\mathcal{Q}} + \lim_{r\to\infty}\frac{1}{32\pi G\,L^2} \,r\, e^{-\tilde{\mathcal{W}}/2} \left( 2 r\, U'+ r^2\, {\phi'}^2 U - 4\, U \right)~,
\end{align}
where $\mu\equiv\frac{1}{\sqrt{16\pi G}}A_\tau(\infty)$ and $\tilde{Q}$ are the chemical potential and the charge density in the dual field theory and we also identified $\mathcal{Q}$ with $\frac{1}{\sqrt{16\pi G}}\tilde{\mathcal Q}$.

Since we are taking into account a homogeneous system, we may expect that the following relation
using the thermodynamic potential density($\omega = -\mathcal{P}$) in (\ref{pressure}):
\begin{align}
\epsilon + \mathcal{P} =& \,\mu\,\tilde{\mathcal Q} + s\, T_H \\
=&  \lim_{r\to\infty}\frac{1}{32\pi G L^2} \,r\, e^{-\tilde{\mathcal{W}}/2} \left( 2 r\, U'+ r^2\, {\phi'}^2 U - 4\, U \right) + \frac{1}{32\pi\,G\,\lambda}\int_{r_H}^\infty dr'\delta_H L_H(r') ~,   \nonumber
\end{align} 
where $\epsilon$ is the internal energy given by one component of the holographic tensor $T_{\tau\tau}$. Furthermore the pressure $\mathcal P$ can be decomposed into the two parts: $\mathcal{P}_{others}+\mathcal{M}\,\mathcal{B}$, where we have used the magnetic field $\mathcal{B}\equiv \sqrt{L}\,H/\sqrt{16\pi\,G}$ as a field theory quantity. Since we don't consider sources except for the chemical potential and the magnetic field, $\mathcal{P}_{others}$ is simply given by the holographic pressure, $ T_{xx}$. The holographic energy-momentum tensor is usually obtained from asymptotic values of fields. Therefore we may claim that the magnetization density is given by
\begin{align}\label{MagnetizationGeneral}
\mathcal M=&\, \frac{1}{32\pi\,G\,\lambda\,\mathcal{B}}\int_{r_h}^\infty dr'\delta_H L_H(r') \nonumber\\=&\, - L^2\,\int_{r_h}^\infty dr\, e^{-\tilde{\mathcal{W}}/2}\left(  \frac{\mathcal{B} Z(\phi)}{L r^2}  + \frac{64 \mathcal{B} \theta^2 W^2(\phi)}{L r^2 Z(\phi)} +\frac{8\,\theta \tilde{\mathcal{Q}} W(\phi)}{\sqrt{L}r^2 Z(\phi)}   \right)~,
\end{align}
where $\tilde{\mathcal{Q}}$ is the charge density $\tilde{\mathcal{Q}} = \sqrt{16\pi G} \mathcal{Q}$ in the dual field theory. By using an appropriate coordinate transformation, it turns out that this expression is same with formula for the magnetization in \cite{Donos:2012yu,Kim:2015wba,Blake:2015ina,Donos:2017mhp}. 
Our conjectured form of the magnetization covers all these cases. Each case has specific forms of $Z(\phi)$ and $W(\phi)$ and depends on counter terms in holographic renormalization. In particular, the authors of Ref. \cite{Donos:2017mhp} focus on the DC transport which can be one of our future directions.

Looking at the formula (\ref{MagnetizationGeneral}), one can notice that something interesting happens. Even though $\mathcal{B}=0$, the magnetization still exists if scalar field has finite expectation value.  This means that $\mathbb{Z}_2$ symmetry of the magnetization is spontaneously broken with the help of the hairy configuration of the scalar field. Existence of this magnetic ordering was already mentioned in \cite{Donos:2012yu} for a specific model. In the next section we will investigate this broken phase in a different model and show that this phenomenon describes spontaneous magnetization and hysteresis behavior.
The resultant spontaneous magnetization is given as follows:
\begin{align}
\mathcal{M}_{f}\equiv\mathcal{M}_{\mathcal{B}=0} = -\, L^{3/2}\int_{r_h}^\infty dr\, e^{-\tilde{\mathcal{W}}/2}\left(  \frac{8\,\theta \tilde{\mathcal{Q}}  W(\phi)}{r^2 Z(\phi)}   \right)~~.
\end{align}
One can see that the sign of $\theta$ determines that of the magnetization. Under proper conditions, the spontaneous magnetization is accompanied by a hysteresis curve which is  a trajectory in $\left(\mathcal{B},\mathcal M\right)$-configuration space. In order to obtain such a curve, we have to consider black brane solutions with a real scalar hair in a numerical method. In the following sections, we will discuss the spontaneous magnetization and the corresponding hysteresis curves in a particular model. Also, we will check whether our expression (\ref{MagnetizationGeneral}) is valid or not in the model.

\subsection{ A Simple Model }\label{AsimpleModel}

In this subsection we consider a specific model which is relevant to a spontaneous magnetization and a hysteresis curve. We start with adopting $Z(\phi)=1$, $V(\phi) = - \frac{2}{L^2} \phi^2$ and $W(\phi) = - \phi^n$. As we discussed, we will consider only $n=1$ as a representative of odd $W(\phi)$ cases. Together with this choice, the real scalar field is supposed to be dual to a dimension 2 operator ($\Delta = 2$) for simplicity.
Then the total action consists of the bulk part and the boundary part as follows\footnote{ $ x^M =( x^\mu , r  ) = (\tau, x^i, r)=( \tau , x, y, r) $} :
\begin{align}\label{explicitModel}
S_{total}= & \,S_B + S_b~, \\
S_B = & \,\frac{1}{16\pi G}\int_{\mathcal M} d^4 x \sqrt{-g}\, \mathcal{L}_{bulk}\nonumber\\
= &\frac{1}{16\pi G}\int_{\mathcal M} d^4 x \sqrt{-g}  \left\{ R + \frac{6}{L^2}    -\frac{1}{4} F^2 - \frac{1}{2} \left(\partial \phi\right)^2    +\frac{1}{L^2}  \phi^2     \right\} \nonumber\\&- \frac{\theta}{16\pi G} \int_{\mathcal M} d^4 x \phi^n \epsilon^{MNPQ} F_{MN} F_{PQ}~, \\
S_{b}=&\,- \frac{1}{16\pi G}\int_{\partial  \mathcal{M} }d^3 x  \sqrt{-\gamma}  \left( 2 K +  \frac{4}{L} +  \frac{\phi^2}{2 L }   \right)~,
\end{align}
where $K$ is the Gibbons-Hawking term which is the tace of extrinsic curvature. The other terms in the boundary action $S_b$ are counter terms for holographic renormalization. The equations of motion for matter fields are given by
\begin{align}\label{FphiEq}
&\nabla_M  \left( F^{MN} +4 \,\theta\,  \phi^n     \frac{1}{ \sqrt{-g}}  \epsilon^{MNPQ} F_{PQ}  \right)=0  \\
& \left( \nabla^2 + \frac{2}{L^2} \right) \phi  - n\,\theta\, \phi^{n-1} \frac{1}{\sqrt{-g}}\epsilon^{MNPQ} F_{MN}F_{PQ} =0~~.
\end{align}
And the Einstein equation is
\begin{align}\label{EinEq}
R_{MN} - \frac{1}{2} g_{MN} \left(   R +\frac{6}{L^2} - \frac{1}{4} F^2 - \frac{1}{2} ( \partial \phi )^2  + \frac{1}{L^2} \phi^2 \right) - \frac{1}{2} F_{MP} {F_N}^P  -\frac{1}{2} \partial_M \phi \partial_N \phi =0 ~. 
\end{align}
In addition, the magnetization formula (\ref{MagnetizationGeneral}) becomes
\begin{align}\label{MagExpl}
\mathcal M = - L\int_{r_h}^\infty dr e^{-\tilde{\mathcal{W}}/2}\left(  \frac{\mathcal{B}}{r^2}  + \frac{64\, \theta^2 \mathcal{B}\,\phi^{2n}}{r^2} -\frac{8\sqrt{L}\,\theta\,\tilde{\mathcal Q}\, \phi^n}{r^2}   \right)~~,
\end{align}
In the next subsection, this formula will be checked by holographic renormalization.

\subsection{Magnetization with Holographic Renormalization}\label{sub2.3}

Now, we will show that our proposal for magnetization (\ref{MagnetizationGeneral}) is correct in the explicit model (\ref{explicitModel}) using holographic renormalization. In order to show the validity, we first need to find the Euclidean on-shell action. We extend similar calculations for the on-shell action in \cite{Hartnoll:2008kx,Kim:2015dna}. The trace of the Einstein Equation (\ref{EinEq}) is given by 
\begin{align}
R = 2 \left(\mathcal{L}_{\text{bulk}}+\frac{\theta  }{\sqrt{-g}} \phi^n F_{M N} F_{P Q} \epsilon ^{M N P Q}\right) + \frac{1}{2}F^2 + \frac{1}{2}(\partial \phi )^2
\end{align}
or
\begin{align}
& 2 g^{x x} G_{x x}+G^{\tau}{}_{\tau}+G^r{}_r + 2 R \\
&~~~~= 2 \left(\mathcal{L}_{bulk}+\frac{\theta }{\sqrt{-g}} \phi ^n F_{M N} F_{P Q} \epsilon ^{M N P Q} \right) + \frac{1}{2}F^2 + \frac{1}{2}(\partial \phi )^2 ~~, \nonumber
\end{align}
where $G_{MN}$ is the Einstein tensor. And one can find $G_{xx}$,
\begin{align}
G_{xx} = \frac{r^2}{2 L^2} \left( \mathcal{L}_{bulk} -R +\frac{\theta }{\sqrt{-g}} \phi ^n F_{M N} F_{P Q} \epsilon ^{M N P Q} \right) +\frac{L^2}{2r^2}H^2
\end{align} 
from $xx$-component of the Einstein equation. Using the above three equations, the $\mathcal{L}_{bulk}$ is given by 
\begin{align}
\mathcal{L}_{bulk}= -G^{\tau}{}_{\tau}-G^r{}_r-\frac{L^4}{r^4} H^2-\frac{\theta }{\sqrt{-g}}\phi ^n F_{M N} F_{P Q} \epsilon ^{M N P Q}~.
\end{align}
To get the Euclidean bulk action, we should take the wick rotation of $\tau$.  The bulk action is given by the following integration:\footnote{Here $\tau=-i\tau_E$ and $S_B^E+S_b^E = - i ( S_B+ S_b)$.}
\begin{align}
S^E_B =& - \int_0^{1/T_H} d\tau_E \int d^2 x \int_{r_h}^{\infty} dr\\
& \left( \frac{(-2r\, U\, e^{ \tilde{-\mathcal{W}}/2  }  )'}{16\pi G L^2} - \mathcal{B}L^2 e^{-\tilde{\mathcal{W}}/2} \left(\frac{\mathcal{B}}{r^2} -\frac{8 \theta \tilde{\mathcal{Q}} {\phi}^n}{r^2}+\frac{64 \theta ^2 \mathcal{B} {\phi}^{2 n}}{r^2} \right)  \right)\nonumber~,
\end{align}
where $\tau_E$ is the Euclidean time and one can notice that the proposed magnetization expression (\ref{MagExpl}) appears in the bulk on-shell action. 

To finish the calculation of the total on-shell action, let us move on to the boundary part. The Euclidean boundary action is given by
\begin{align}
S_b^E &=\lim_{r\to\infty} \frac{1}{16\pi G} \int_0^{1/T_H} d\tau_E \int d^2 x  \sqrt{-\gamma}  \left( 2 K +  \frac{4}{L} +  \frac{\phi^2}{2 L }   \right) \\
&=\,\frac{1}{16\pi G L^2} \frac{1}{T_H}\int d^2 x\lim_{r\to\infty}\,r\, e^{-\tilde{\mathcal{W}}/2}\left(r \sqrt{U} \left(\frac{4}{L}+\frac{\phi^2}{2 L}\right)-r\, U' - U (4-r\, \tilde{\mathcal{W}}') \right)~~.
\end{align}
By plugging the asymptotic behavior of the fields in Appendix, finally we obtained the on-shell action as follows:
\begin{align}\label{E_on-shell}
S_B^E + S_b^E = \frac{\mathcal{V}}{T_H} \,\omega = - \frac{\mathcal{V}}{T_H} \left( \frac{\epsilon}{2} + \mathcal{M}\,\mathcal{B} \right),
\end{align}
where $\mathcal M$ is same with (\ref{MagExpl}). Also, one can notice that $\epsilon/2$ is equivalent to the internal pressure $ \epsilon/2= T_{xx} = 1/2\, T_{\tau\tau}$ which is derived in Appendix. This relation supports the fact that the proposed form of magnetization (\ref{MagnetizationGeneral}) is correct.

\section{Spontaneous Magnetization and Hysteresis Curve}\label{sec3}

In this section we analyze the spontaneous magnetization and hysteresis curve in the explicit model (\ref{explicitModel}) by numerical methods. For numerical calculations, we will take  $16\pi\, G = 1$, $L=1$ and $r_h=1$. In contrast to usual discussions in holographic models, solutions with lager free energy are also important in our discussion. We will explain the role of such unstable solutions in more details in the following subsections.

\subsection{Spontaneous Magnetization}

We concentrate on the case with vanishing magnetic field in this subsection. Then the equations of motion  are reduced to those without the magnetic field.
In high temperature, the AdS Reisner-N$\ddot{o}$rdstrom(RN) Black brane is the preferable solution dual to normal phase. On the contrary, the RN black brane becomes unstable and changes to a hairy black brane dual to an ordered phase in low temperature. The hairy scalar describes the VEV of a dimension 2 operator $\mathcal O$. As we discussed, this hairy scalar contributes to the magnetization (\ref{MagExpl}) even in the absence of the magnetic field. Therefore this holographic model can describe a phase transition from normal phase to spontaneous magnetization phase. Here the sign of the magnetization is determined by the sign of $\theta$ and we have to fix it with a suitable value. For an odd $n$, both signs of magnetization are possible with a given $\theta$. On the other hand the sign of magnetization is fixed for an even $n$. In the latter case the direction of magnetization is always same for both positive and negative scalar condensations. This is the reason why we concentrate on the $n=1$ case because we need the both signs of magnetization to construct magnetic hysteresis.

\begin{figure}[ht!] 
    \subfigure[ ]
    {\includegraphics[width=6.5cm]{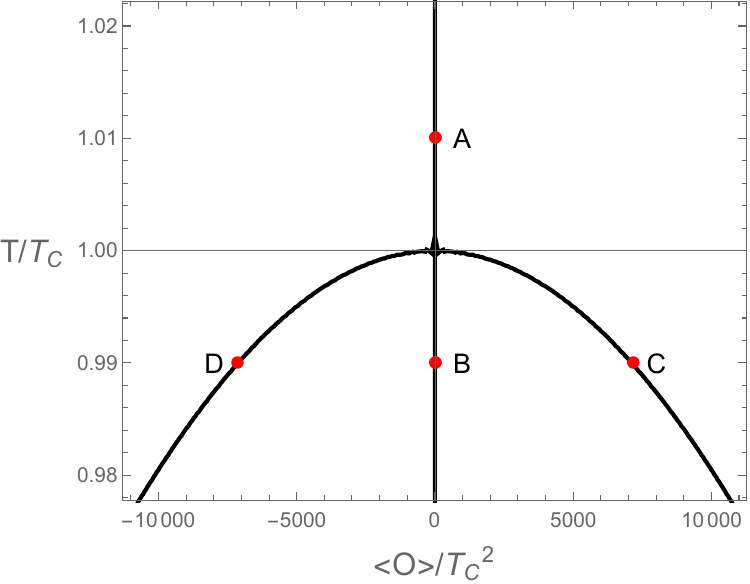}  }
    \hskip1cm
    \subfigure[ ]
    {\includegraphics[width=7.5cm]{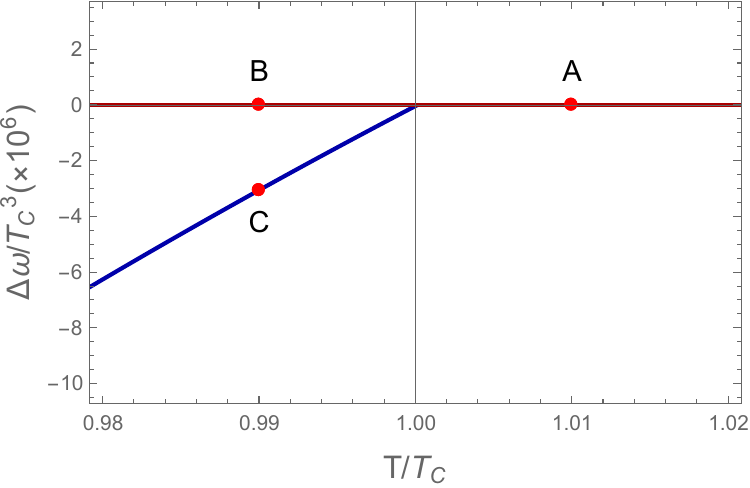}  }\\

    \subfigure[ ]
    {\includegraphics[width=7cm]{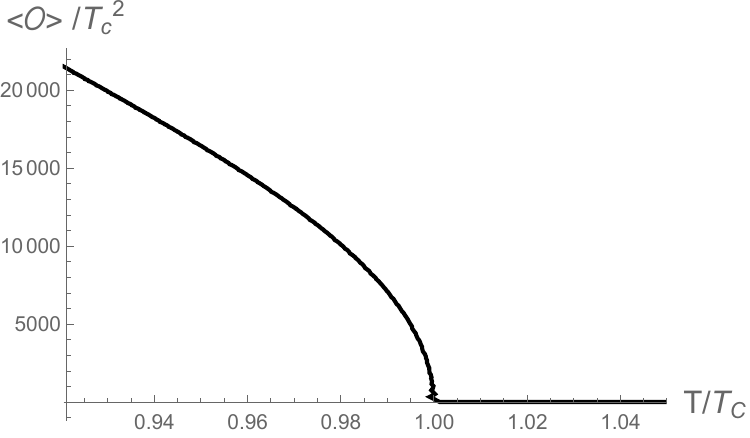}  } 
   \hskip1cm
    \subfigure[ ]
    {\includegraphics[width=7cm]{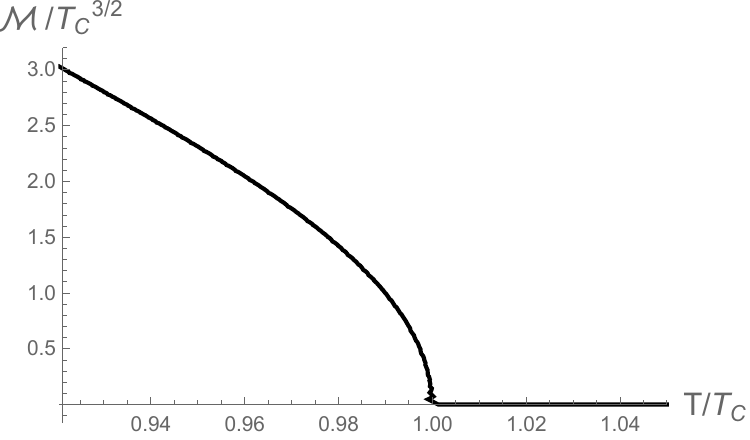}  }

    \caption{(a) The set of black brane solutions satisfying zero source condition (b) The  difference of free energy of RN black branes(red) and hair black branes(blue). (c) Temperature dependence of the absolute value of the condensation. (d) Temperature dependence of the absolute value of the magnetization, where we took $\theta= - 0.2$ for numerical calculation.
} \label{fig:Hzero}
\end{figure}

Also, if we allow only solutions without source of scalar condensation, i.e, $J_{\mathcal{O}}=0$ in (\ref{AsympS}), an instability of the RN black brane happens when $T_H=T_c$. 
Figure \ref{fig:Hzero} (a) shows a set of black brane solutions satisfying the sourceless condition. Above this critical temperature, the only solution is that of zero condensation which is nothing but the RN black brane. On the other hand, two more kinds of hairy black brane solutions show up below $T_c$(solid curve connecting C and D). They describe positive and negative values of $\left<\mathcal{O}\right>$ naturally coming from the $\mathbb{Z}_2$($\phi \leftrightarrow -\phi$) symmetry in the action (\ref{explicitModel}).

Figure \ref{fig:Hzero} (b) presents free energy difference between RN black brane solutions and the hairy ones. Red solid line denotes the free energy of the RN black branes as a reference and blue line corresponds to the free energy of the hairy black branes. Below $T_c$, the free energy of the hairy solutions is lower than RN solutions. Therefore there is a second order phase transition from RN to hairy black brane solution at $T=T_c$.

We plot the condensation diagram in Figure \ref{fig:Hzero} (c). Since the integrand of magnetization density is proportional to $\left<\mathcal{O}\right>$ near critical temperature, the magnetization shows similar behavior as shown in Figure \ref{fig:Hzero} (d). Since the magnetization and the condensation are proportional to $\sqrt{T_c-T}$ near critical temperature, one can notice that the transition is a second order phase transition\footnote{Figure \ref{fig:Hzero} (b) looks like a first order phase transition. This is coming from numerical difficulty due to very low critical temperature. We checked that the deviation between RN black brane and hairy black brane has smooth behavior by considering well-scaled physical quantities.  }. Also, the charge density is essential to produce the magnetization. Thus it turns out that the charge carrier carries magnetic moment and there exists an interaction between the charge operator $\tilde{{\mathcal Q}}$ and the dimension 2 operator $\mathcal O$ in the dual field theory.

The scalar condensation comes from the spontaneous symmetry breaking. In the dual gravity picture, this corresponds to advent of normalizable hairy scalar around the black brane solutions. The holographic superconductor model admits a complex scalar hair which is associated with spontaneous breaking of continuous $U(1)$ symmetry \cite{Hartnoll:2008vx}. On the other hand, for this real scalar, the associated symmetry is now $\mathbb{Z}_2$ symmetry. Therefore, these phenomena can be understood as a spontaneous symmetry breaking of $\mathbb{Z}_2$ symmetry of the scalar operator and the magnetization due to the axion-like coupling $\int \phi\, F\wedge F$ in the absence of magnetic field. In a field theoretic description, the spontaneous symmetry breaking can be illustrated through a suitable Landau-Ginzberg potential. Although such a description can not have all the information of holographic model, it may give an intuition about what's going on during the phase transition. So we provide plausible explanation using a Landau-Ginzburg potential in Appendix B.

\subsection{Hysteresis Curve}\label{hysteresis}

In the previous subsections, we showed that a holographic model can describe the spontaneous magnetization and the hairy black brane solution is dual to the broken phase. As we mentioned in the introduction, one of important properties of magnetic system is the hysteresis curve. We now show how to construct the hysteresis curves. For this purpose,  we first find the hairy black hole solutions with the non-vanishing magnetic field using a numerical method. We impose the boundary condition $J_{\mathcal{O}}=0$ for the scalar field $\phi$ in (\ref{AsympS}) to simplify the problem. 

\begin{figure}[ht!]
\centering
    {\includegraphics[width=10cm]{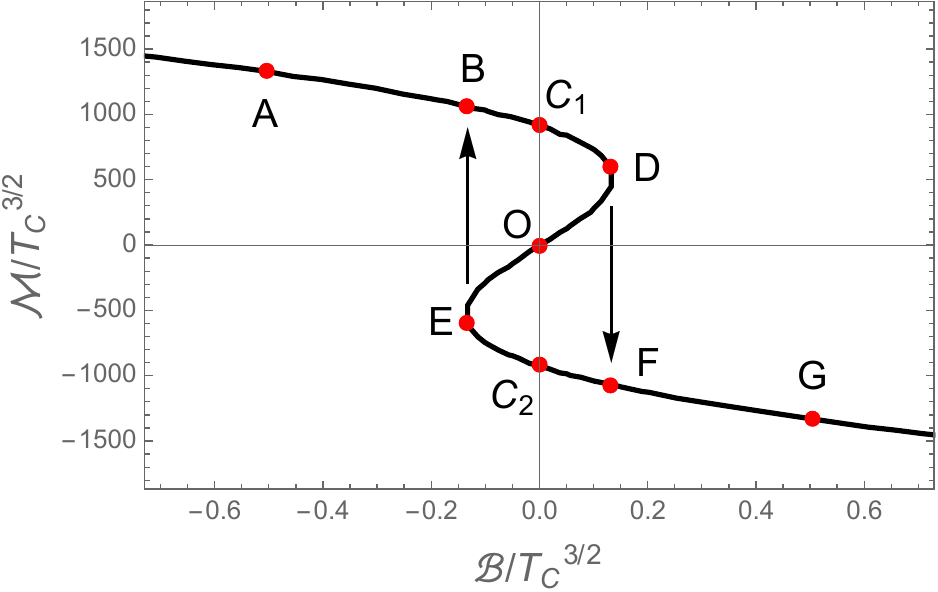}  }     
\caption{External magnetic field dependence of the magnetization.}
      \label{fig:hys01}
\end{figure}

\begin{figure}[ht!]
\centering
    \subfigure[ ]
    {\includegraphics[width=6.5cm]{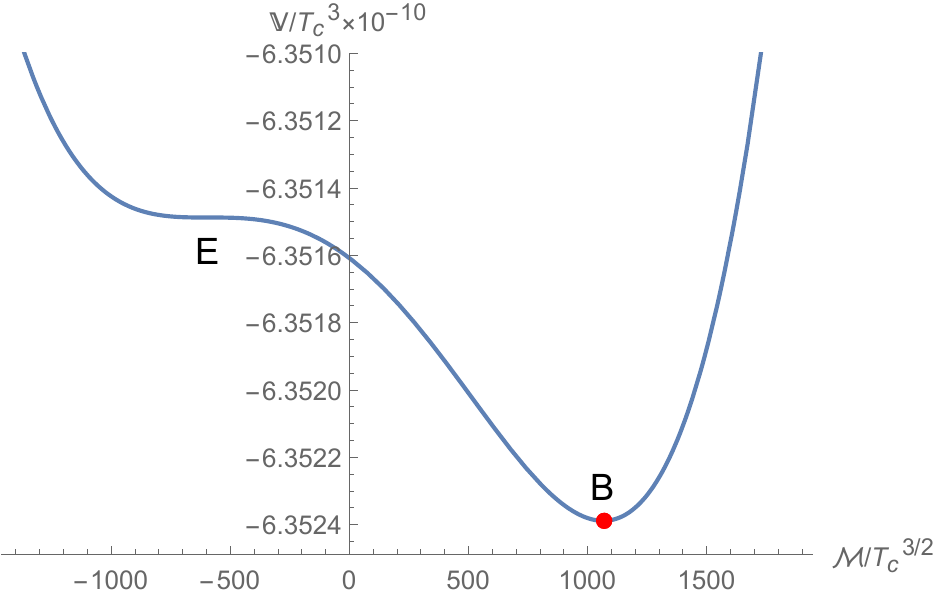}  }
    \hskip1cm
        \subfigure[ ]
    {\includegraphics[width=6.5cm]{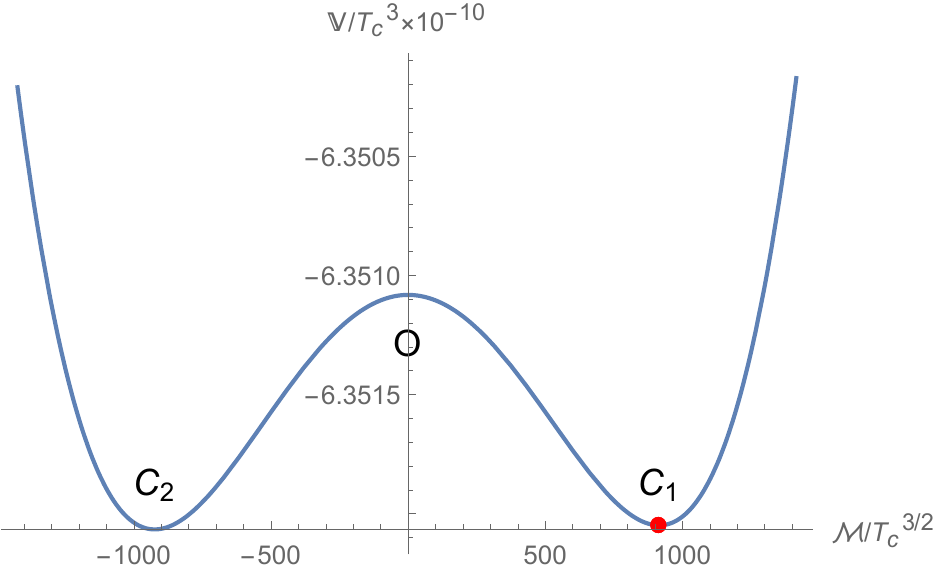}  }\\ 
     \subfigure[ ]
    {\includegraphics[width=6.5cm]{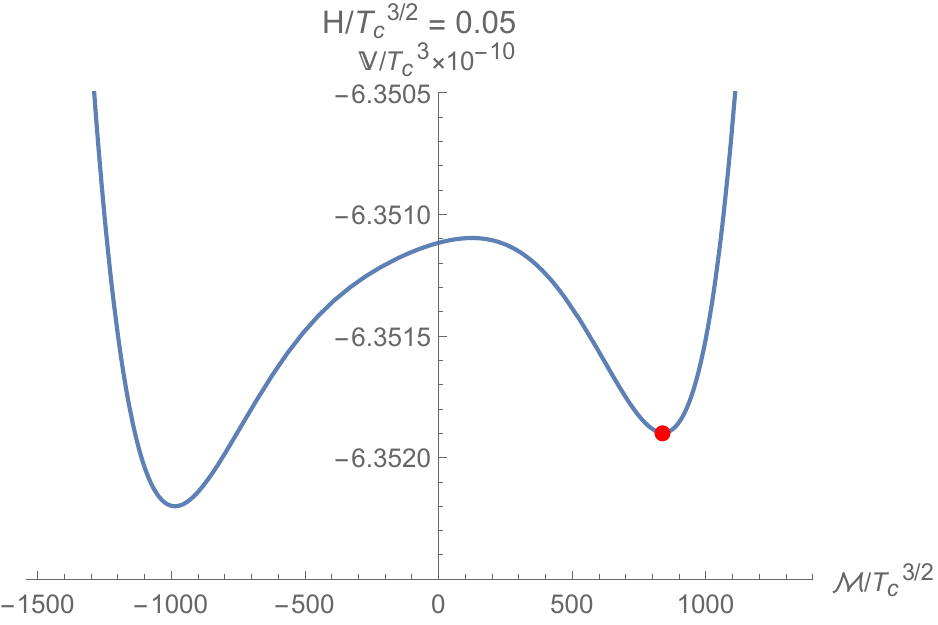}  }
    \hskip1cm
    \subfigure[ ]
    {\includegraphics[width=6.5cm]{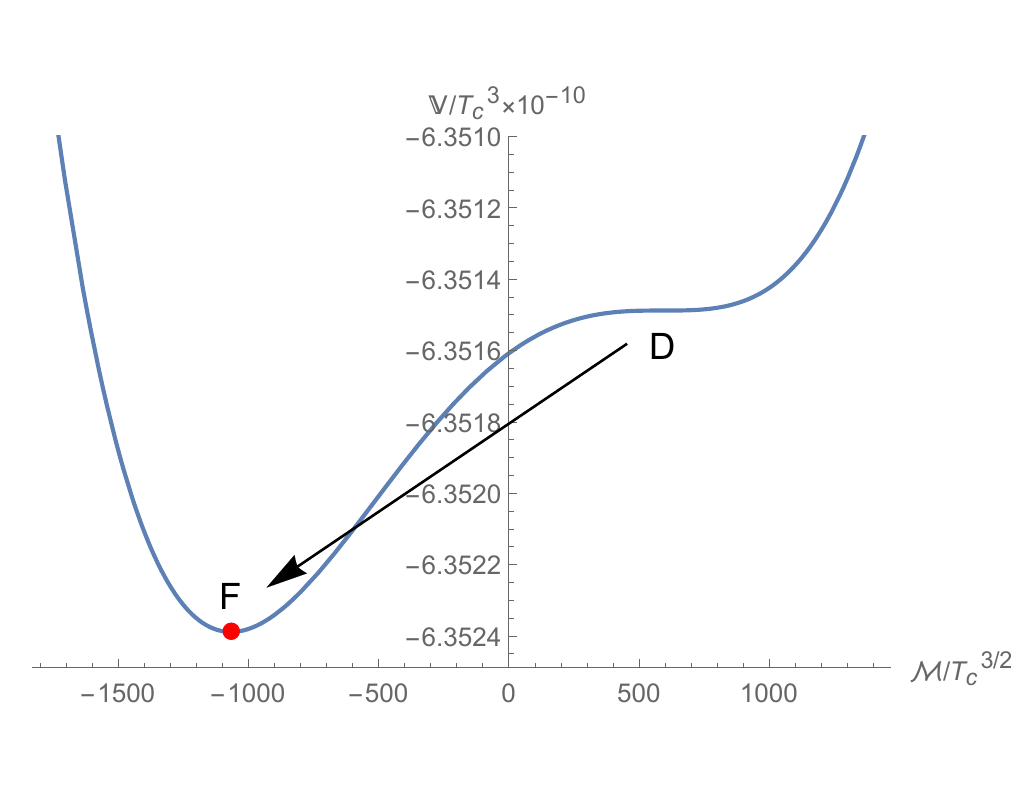}  }
\caption{Magnetization dependence of the effective potential for given magnetic field corresponding to Figure \ref{fig:hys01}. The potentials (a) (b) (c) and (d) are given upto $\mathcal{M}^6$-\,order.}
      \label{fig:MF}
\end{figure}

Using the solutions below the critical temperature, we display the  relation of external magnetic field and magnetization in Figure \ref{fig:hys01}: there exist three solutions  for small magnetic field. The solutions on (D-O-E) line have higher free energy than those on (B-$C_1$-D)  and (E-$C_2$-F) lines at the same magnetic field. 
Just for convenience of explanation, we may imagine a Landau-Ginzburg type potential which is not essential way to understand these phenomena though. See Figure \ref{fig:MF}.
With the help of L-G description, the (D-O-E) line represent unstable extrema of the free energy potential, while the other (B-$C_1$-D)  and (E-$C_2$-F) lines describe the stable and metastable extrema of the potential. If the black brane solution changes following the local minimum of the potential under slowly varying magnetic field, such process can give rise to a magnetic hysteresis.

Let us describe how the potential changes in more detail. We start with A in Figure \ref{fig:hys01} and increase the external magnetic field slowly. Until the magnetic field reaches B, there is a single minimum and it should be located at the minimum of the potential, see Figure  \ref{fig:MF} (a). When the magnetic field becomes zero, there are three solutions. Two of them are at the minima   and the other one is at the local maximum(RN black brane), see Figure \ref{fig:MF} (b).  Further variation of the magnetic field gives rise to an asymmetric potential. If the thermal fluctuation and the changing rate of the magnetic field are very small, then the system can be placed on a metastable point in the potential. See Figure \ref{fig:MF} (c). When the magnetic field reaches D in Figure \ref{fig:hys01}, the local minimum where the solution is located becomes unstable, Figure \ref{fig:MF} (d). Then, the solution cannot stay there and moves to F and this change looks like a sudden jump of the magnetization. The solution will stay there as the magnetic field increases further. If we start from large magnetic field and decrease it, then the process will be reversed.

The potentials in Figure \ref{fig:MF} are obtained by interpolating three extremal values. Letting $f=\mathbb{V}/T_c^3$ and $m=\mathcal{M}/T_c^{3/2}$, we have 6 conditions which consist of free energy $f(m_i)$ and extremal condition $f'(m_i)=0$ corresponding to three hairy solutions for a given small $\mathcal{B}$ in Figure \ref{fig:hys01}. Here the index $i$ is running from 1 to 3 and $m_i$ denotes magnetization of three hairy black branes. From the data, one can find $f(m)$ as polynomials of $m$. In order to make W-type potential, we add $m^6$ term by hand. So these potentials $f(m)$ are good approximation up to $m^5$ order for small $m$. Our result describing magnetic hysteresis is reminiscent of the bistable model accompanied by the Barkhausen effect in condensed matter physics. See \cite{Bertotti1998} for a text book.

While generating the hysteresis curve, we need to pay attention to one possibility of a transition from the local minimum to the global minimum. In order to investigate this possibility, we have to consider time evolution of the background\footnote{See \cite{Evans:2010xs} for an example of time dependent process in a non-equilbrium physics}. This is a challenging issue and beyond present discussion. In this work we restrict our study only for static configurations. In other words, extremely slow variation of magnetic field is assumed for generating hysteresis.

\begin{figure}[ht!]
\centering
    \subfigure[ ]
    {\includegraphics[width=7cm]{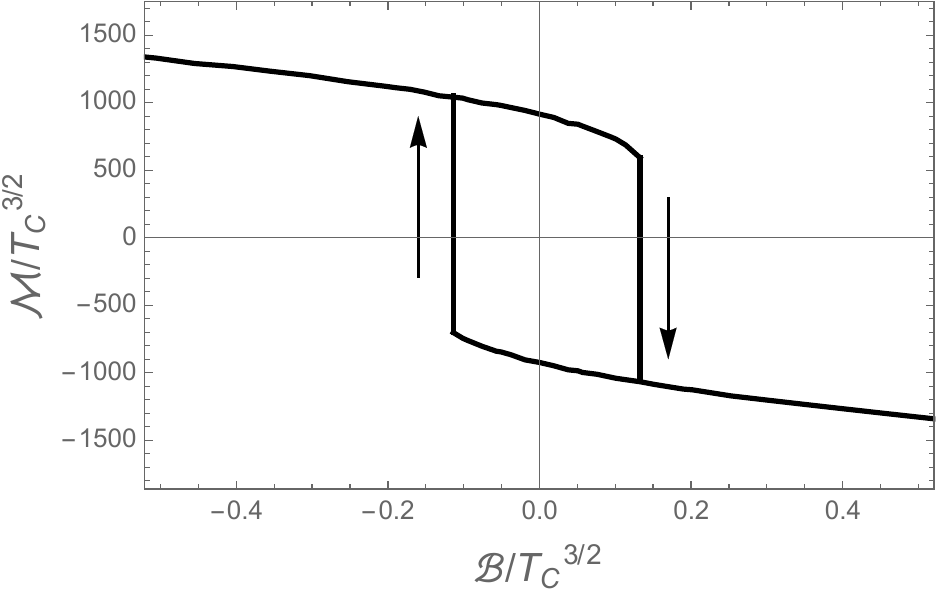}  }
    \hskip1cm
    \subfigure[ ]
    {\includegraphics[width=7cm]{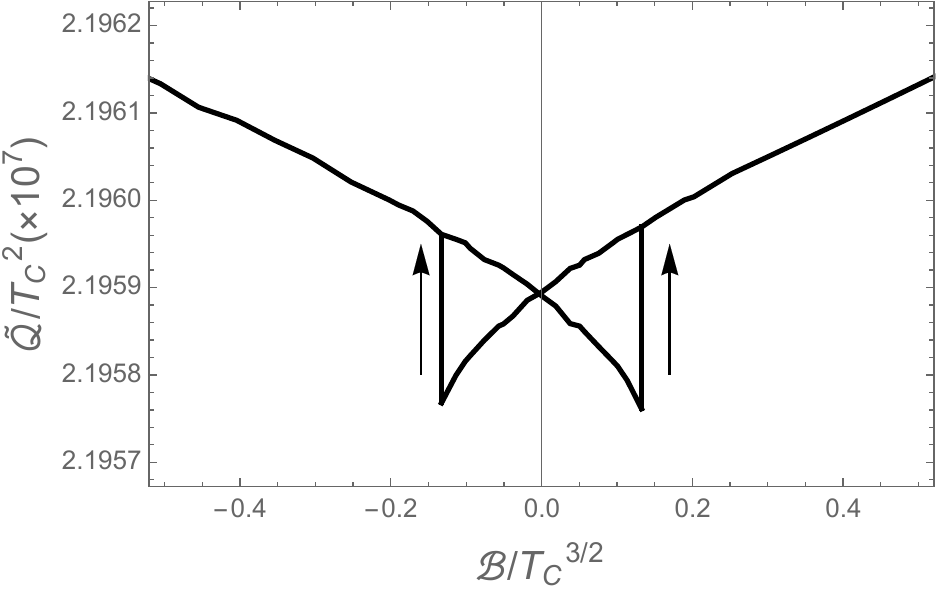}  }
     
    \caption{(a) Hysteresis curve of the magnetization. (b) Hysteresis curve of the charge density. }
      \label{fig:hys02}
\end{figure}

The resulting process of the magnetization hysteresis is drawn in Figure \ref{fig:hys02} (a). Figure \ref{fig:hys02} (b) shows hysteresis behavior of the charge density for (a). The charge carrier density is decreasing for both direction and suddenly increase when the magnetization changes.

\begin{figure}[ht!]
\centering
    {\includegraphics[width=7cm]{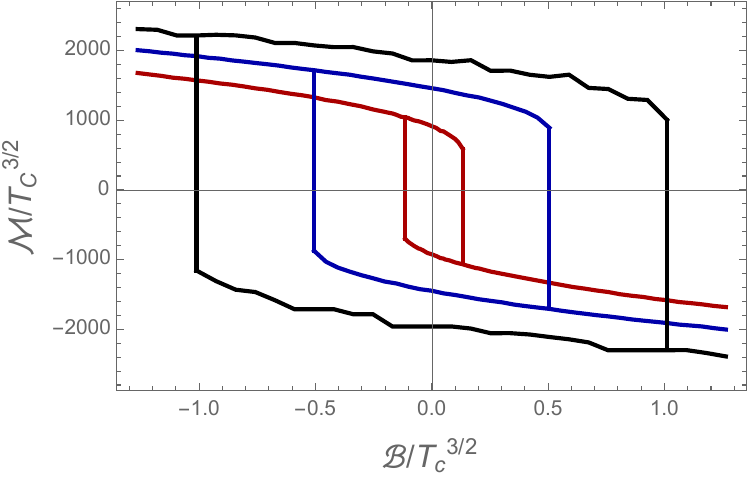}  }
     
    \caption{Temperature dependence of the hysteresis curve with $T/T_C =$ 0.98(red), 0.96(blue), 0.93(black). }
      \label{fig:hys03}
\end{figure}

In Figure \ref{fig:hys02}, non-zero values of the magnetization cause the Hall effect and it gives nontrivial Hall conductivity($\sigma_{xy}$) or Hall resistance($\rho_{xy}$). We expect that there also be hysteresis behavior to the Hall resistance. On the other hand,  the charge carrier density $\mathcal{Q}$ would be proportional to the longitudinal conductivity $\sigma_{xx}$. The results are qualitatively same to the transport on the surface state of the topological insulator. See Figure 3 and 4 in \cite{Bao2013}. In addition, Figure \ref{fig:hys03} shows temperature dependence of the hysteresis curve. As temperature decreases, the size of hysteresis curve increases. It is natural that the expectation value of the scalar is bigger at lower temperature and so is the magnetization.

\section{Discussion}\label{sec5}

In this paper, we have studied the magnetic properties of the strongly interacting system using gauge/gravity duality. To do this, we first defined the magnetization of the boundary theory in terms of the bulk data using the `scaling symmetry' trick. Then, we applied it to a specific model and confirmed that the formula  (\ref{MagnetizationGeneral}) agreed with the magnetization obtained from the holographic renormalization and the corresponding magnetization satisfied thermodynamics laws.
In our model, the magnetization comes from the condensation of the scalar field due to the spontaneous symmetry(discrete $\mathbb{Z}_2$) breaking. By introducing a coupling between magnetic field and the real scalar, we showed that a spontaneous magnetization occurs in our model.

Also, we analyzed the field theory systems with the magnetic field dual to the magnetic hairy black branes. Using the black brane solutions, we showed that the set of successive solutions with varying magnetic field described the magnetic hysteresis. The hysteresis curve we obtained can be compared to models in condensed matter physics\cite{Bertotti1998}. In addition, we also obtained hysteresis behavior of the charge density.

The magnetization hysteresis can be generated by slowly varying magnetic field. In this hysteresis process, the time scale of the variation is important. So we assume very slow change of the system. More detailed discussion would be interesting as a future direction. In addition to the issues, our resulting hysteresis has rapidly varying shapes and the fragmentation can make the shapes more smoothly. However, we did not consider the fragmentation of magnetization because the corresponding black brane solution is not easy to obtain. Thus we leave it as another future study.

As a final comment, it is expected that the magnetization and the charge carrier density are proportional to the Hall resistivity and longitudinal conductivity, respectively. Therefore, there exists hysteresis behavior of these quantities~\cite{Bao2013,Chang2013,Barzola2012}  . A study on this subject will be reported soon \cite{KimSeo}.

\section*{Appdendix}

\subsection*{A. Holographic Renormalization}

In this section we provide expression of boundary tensors such as energy-momentum tensor, current and condensation of scalar operator dual to bulk scalar field $\phi$ in the subsection \ref{AsimpleModel}. In order to get these formulas using holographic renormalization, we introduce ADM decomposition without the shift vector. Thus the metric (\ref{ansatz0}) can be decomposed as follows:
\begin{align}
ds^2= \gamma_{\mu\nu} dx^\mu dx^\nu + N^2 dr^2~~,
\end{align}
where $\gamma_{\mu\nu}= \text{diag}\left( -U(r)\,e^{\tilde{\mathcal W}(r)},\frac{r^2}{L^2},\frac{r^2}{L^2} \right)$ and $N = 1/\sqrt{U(r)}$. Then the extrinsic curvature tensor and the Gibbons-Hawking term $K$ are given by $K_{\mu\nu}=- \frac{1}{2N}\gamma'_{\m\nu}$ and $\gamma^{\mu\nu}K_{\mu\nu}$, respectively. A boundary tensor is usually given by variations of the total action $S_B + S_b$ with respect to corresponding sources. In this model, they are given by
\begin{align}\label{asympto01}
&\left<T_{\mu\nu}\right> = \frac{1}{16\pi G}\lim_{r\to\infty} 2 \frac{r}{L} \left[ K_{\mu\nu}- K \gamma_{\mu\nu} - \left( \frac{\phi^2}{4 L} +\frac{2}{L}  \right)\gamma_{\mu\nu}  \right]\\
&\left<J^\mu\right> =  \frac{-1}{\sqrt{16\pi G}}\lim_{r\to\infty} \frac{\sqrt{-g}}{\sqrt{-\gamma}}F^{r\mu} \\
&\left<\mathcal{O}\right> =\frac{L}{\sqrt{16\pi G}} \lim_{r\to\infty} \frac{1}{r} \left( -\sqrt{-g}\,\nabla^r \phi - \sqrt{-\gamma}\, \phi/L \right)
\end{align}

On the other hand, the asymptotic behavior of the bulk geometry is given by
\begin{align}
&\tilde{\mathcal{W}}(r) \sim 4\pi G \left(J_{\mathcal{O}}^2 \left(\frac{L}{r}\right)^2 + \frac{8\, L\, J_{\mathcal O}\tilde{\mathcal{O}}}{3}\left(\frac{L}{r}\right)^3+\ldots \right)\\
&U(r) \sim \left(\frac{r}{L}\right)^2 +4\pi G\left( J_{\mathcal{O}}^2 - 2 \, L\, \epsilon \left(\frac{L}{r}\right) +\ldots\right)\\
&\phi(r)\sim \sqrt{16\pi G}\left(J_\mathcal{O}\left(\frac{L}{r}\right) +   L \tilde{\mathcal{O}}\left(\frac{L}{r}\right)^2 + \ldots \right)~~,\label{AsympS}\\
&A_\tau (r) \sim \sqrt{16\pi G}\left(\mu - L\, \tilde{\mathcal{Q}} \left(\frac{L}{r}\right)+\ldots\right)~,
\end{align}
where we are considering $J_{\mathcal{O}}=0$ case since we don't want to introduce another source except for the chemical potential and the magnetic field\footnote{ One can easily extend the case with a non-vanishing $J_{\mathcal{O}}$.}. Then, we can obtain the boundary tensors as follows:
\begin{align}
\left<T_{\mu\nu}\right> = \left(\begin{array}{ccc}
\epsilon & 0 & 0 \\ 
0 & \epsilon/2 & 0 \\ 
0 & 0 & \epsilon/2
\end{array}\right)~,~\left<J^\mu\right> =( \tilde{\mathcal Q},0,0)~,~\left< \mathcal{O}\right> = \tilde{\mathcal O}\,.
\end{align}

\subsection*{B. Spontaneous Magnetization using Landau-Ginzburg potential}

 This phase transition discussed in the main text can be explained by a Landau-Ginzburg type theory whose effective field is a real scalar field. We denote this field by $\Phi$. Therefore, the potential of the Landau-Ginzburg model governs this phase transition. The typical shape of potential in the broken phase should admit $\mathbb{Z}_2$ symmetry so it has two global minima. At these minima, the field has values, $\Phi=\pm\,\Phi_0$ and free energy of the system is given by the on-shell action of hairy black branes.

\begin{figure}[ht!]
\centering
    \subfigure[ ]
    {\includegraphics[width=7.8cm]{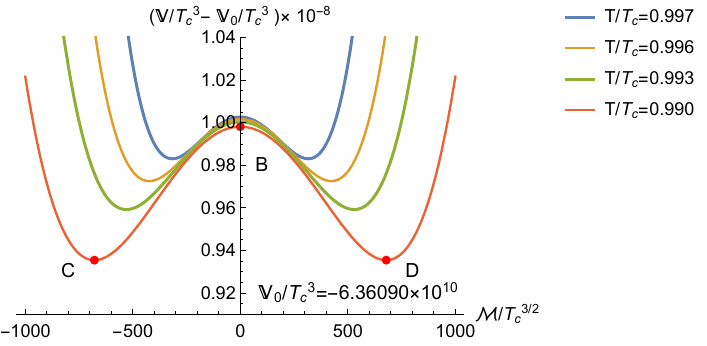}  }
    \hskip0cm
    \subfigure[ ]
    {\includegraphics[width=7.8cm]{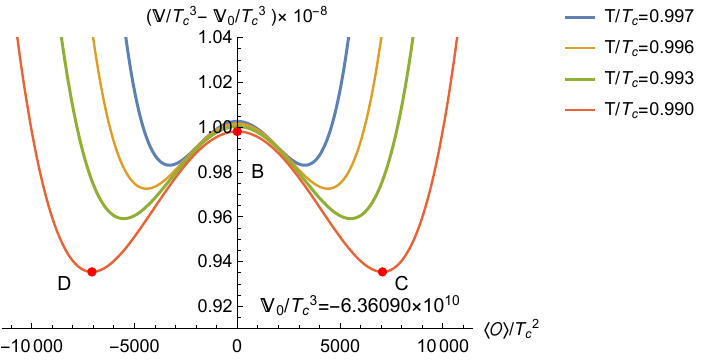}  }
    \hskip0.2cm

    \caption{The effective potential in terms of (a) the magnetization, (b) the condensation. The point B, C and D correspond to the same points in Figure \ref{fig:Hzero} (a) and (b).}
      \label{fig:Cond}
\end{figure}

Now, we construct the Landau-Ginzburg type effective potentials using the data in Figure \ref{fig:Hzero}. One can naturally deduce that the free energy density of the RN black brane solution, such as B in Figure \ref{fig:Hzero}, corresponds to the unstable extremum due to $\mathbb{Z}_2$ symmetry and vanishing condensation. This is quite reasonable because the RN black branes are solutions of the dual gravity system and the extrema of the potentials are also solutions of the corresponding effective theory. In addition to this observation, we may assume that the condensation $\left< \mathcal{O} \right>$ and the magnetization $\mathcal{M}$ are very small near the critical temperature. Then the effective potential of free energy can be approximated by a polynomial of $\mathcal{O}$ or $\mathcal{M}$. The suggested form of the potential is given as follows
\begin{align}\label{V1}
\mathbb{V}(\Phi) = \omega_{RN} + \frac{\omega_{RN}-\omega_{HB}}{\Phi_0^4}  \Phi^2   \left(\Phi^2-2 \Phi_0^2\right) +  \mathcal{O}\left(\Phi^6 \right)~~, 
\end{align}  
where $\Phi$ is the effective field which in our case denotes the magnetization $\mathcal{M}$ or $\mathcal{O}$ and $\Phi_0$ denotes $\mathcal{M}$ or $\left<\mathcal{O}\right>$ of the hairy black branes, respectively. In addition, $\omega_{RN(HB)}$ is the free energy density of the RN black brane (Hairy black brane). Here, we interestingly point out that the parameters of the Landau-Ginzburg potential can be obtained by the dual geometry data.
We plot the effective potentials in Figure  \ref{fig:Cond}. The figure shows how the potential changes as the temperature decreases. Here one can notice that the on-shell action of RN black brane appears as a parameter of the potential. In the following explanation, we will describe that this plays an important role even in the broken phase.

Now, we may consider the fluctuation of $\Phi$ near the local minima. This fluctuation corresponds to an effective quanta in the broken phase. In this phase, the mass of quasi-particles governs major characteristic of the effective theory. We can easily obtain the mass of the quasi particle as follows:
\begin{align}
m_{eff}^2 =8\,\frac{\omega_{RN}-\omega_{HB}}{\Phi_0^2}.
\end{align}
Then, the low energy effective potential for the broken phase up to quartic order is as follows:
\begin{align}
\mathbb{V}_{BP} \sim \frac{1}{2}\omega_{HB} +\frac{1}{2} m^2_{eff}\delta\Phi^2 + \frac{\lambda_3}{3!} \delta\Phi^3  +  \frac{\lambda_4}{4!} \delta\Phi^4~~,
\end{align}
where $\delta\Phi$ is the fluctuation near $\Phi=\Phi_0$ and
\begin{align}
\lambda_3 = 4!\, \frac{\omega_{RN}-\omega_{HB}}{\Phi_0^3} ~~\text{and}~~ \lambda_4=4!\, \frac{\omega_{RN}-\omega_{HB}}{\Phi_0^4} \,.
\end{align}
These couplings describe the qubic and quartic interactions among the quasi particles. Of course, all the mass and couplings are also determined in terms of data from the bulk geometry. When $\Phi=\mathcal{M}$, $\Phi$ describes the fluctuation of the magnetization. This quasi-particle can be interpreted as the magnon whose mass and self-couplings are given by
\begin{align}
m_{magnon}^2 = 8\,\frac{\omega_{RN}-\omega_{HB}}{\left<\mathcal{M}\right>^2 }~,\lambda_3 = 4!\, \frac{\omega_{RN}-\omega_{HB}}{\left<\mathcal{M}\right>^3} ~, \lambda_4=4!\, \frac{\omega_{RN}-\omega_{HB}}{\left<\mathcal{M}\right>^4}\,,
\end{align} 
where $\left<\mathcal{M}\right>$ is obtained by the black brane corresponding D in Figure \ref{fig:Cond}.

\section*{Acknowledgments}
K.K. Kim thanks Koji Hashimoto and Seung-Hyun Chun for helpful comments. The work of K.K. Kim is supported by the faculty research fund of Sejong University in 2018. YS is supported by Basic Science Research Program through NRF grant No. NRF-2016R1D1A1B03931443 and NRF-2019R1I1A1A01057998. SJS is supported by Mid-career Researcher Program through the National Research Foundation of Korea grant No. NRF-2016R1A2B3007687. 
The work of K.-Y. Kim was supported by Basic Science Research Program through the National Research Foundation of Korea(NRF) funded by the Ministry of Science, ICT $\&$ Future Planning(NRF- 2017R1A2B4004810) and GIST Research Institute(GRI) grant funded by the GIST in 2019. We acknowledge the hospitality at APCTP where part of this work was done.


\end{document}